\documentclass[12pt]{article}
\usepackage{amsmath}
\usepackage{amssymb}
\usepackage{graphicx}
\usepackage{float}
\usepackage{indentfirst}
\usepackage{mathrsfs}
\usepackage{dsfont}
\usepackage{pifont}

\usepackage{setspace}

\usepackage[labelfont=bf,labelsep=period]{caption}

\usepackage[numbers,sort&compress]{natbib}

\newtheorem{proposition}{Proposition}

\newtheorem{conjecture}{Conjecture}

\title{An Effective Way of Characterizing the Quantum Nonlocality}

\author
{Ma-Cheng Yang, Jun-Li Li and Cong-Feng Qiao$\footnote{qiaocf@ucas.ac.cn}$ \\ [0.2cm]
\normalsize{School of Physical Sciences, University of Chinese Academy of Sciences,} \\
\normalsize{YuQuan Road 19A, Beijing 100049, China}\\ [2pt]
\normalsize{Key Laboratory of Vacuum Physics, CAS, Beijing 100049, China} \\ [3mm]
}

\date{}

\begin{document}
\baselineskip24pt \maketitle
\begin{abstract}\doublespacing
Nonlocality is a distinctive feature of quantum theory, which has been extensively studied for decades. It is found that the uncertainty principle determines the nonlocality of quantum mechanics. Here we show that various degrees of nonlocalities in correlated system can be characterized by the generalized uncertainty principle, by which the complementarity is attributed to the mutual dependence of observables. Concrete examples for different kinds of non-classical phenomena pertaining to different orders of dependence are presented. We obtain the third order ``skewness nonlocality'', and find that the Bell nonlocality turns out to be merely the second order ``variance nonlocality'' and the forth order dependence contains the commutator squares, which hence is related to the quantum contextuality. More applications of the generalized uncertainty principle are expected.
\end{abstract}

\newpage

\section{Introduction}

In classical physics, observables are represented successfully by real numbers, or something composed of real numbers, e.g. inertia tensor and resistance in Ohm's Law as the direct product of other two physical quantities. And, naturally, it is implicitly assumed that the properties of real numbers are hold by observables. When confronted with the microworld, Heisenberg questioned the assumption by dint of a Gedanken experiment where the canonically conjugated quantities, $x$ and $p$, can only be determined simultaneously with certain indeterminacy \cite{Heisenberg-1927}. Soon afterwards, people realized that the uncertainty relations for incompatible observables are in fact the destined results of quantum mechanics (QM). Unsatisfied with the quantum indeterminacy, Einstein with his collaborators exemplified the "incompleteness" of QM via an entangled bipartite system, viz. the renowned EPR paradox \cite{EPR}, by which and from then on, the inherent nature of QM nonlocality was formally in the spotlight.

To keep on taking the classical recognition on reality and locality, people attempt to construct various models to mimic the QM results with local hidden variables. In order to distinguish the QM from local hidden variable theory(LHVT), Bell put forward a set of inequalities by which all LHVTs should abide \cite{Bell1964}, while quantum theory does not. Among the various Bell inequalities (BIs), one of the most outstnading ones is the Clauser-Horne-Shimony-Holt (CHSH) inequality \cite{CHSH}
\begin{align}
\left| E(X,Y)-E(X,Y') + E(X',Y)+E(X',Y') \right| \leq 2 \; . \label{CHSH-up2}
\end{align}
Here the left four terms denote the correlation functions of observables $X$, $X'$ and $Y$, $Y'$ in a bipartite qubit system. In quantum theory, the left hand side of relation (\ref{CHSH-up2}) may reach $2\sqrt{2}$, breaking the lower bound of 2 \cite{Cirelson}.

A heuristic question regarding the inequality violation may arise: why the quantum limit is $2\sqrt{2}$, but rather not more \cite{PR-axiom}? We know that the Quantum correlations and relativistic causality do not uniquely
define quantum physics. Theories possess the same features with QM but even stronger correlations which break the quantum limit may exist, e.g., communication complexity \cite{Com-complex, Com-complex-RMP} and information causality \cite{Inf-causality}, however, the physical meanings there are still vague \cite{PR-CL}. To determine the fundamental axioms of QM, indirectly, people seek and test the principles beyond the QM. Moreover, Kochen-Specker (KS) contextuality \cite{KS-noncotext1} is known to be a logically independent non-classical concept compared with the Bell nonlocality \cite{Bell-KS}. In the literature, there exist some inequalities witnessing the contextuality \cite{KS-inequality}, but the corresponding theoretical bases still need further investigations \cite{Bell-KS-In}. Considering that the non-classical natures of Bell nonlocality and KS contextulity are potential resources for quantum secure communication \cite{Secure-Ekert} and quantum computation \cite{Contex-QP}, it is tempting to think whether there are some other yet unknown quantum nonlocal phenomena or not. And, if yes, what are the criteria by which various nonlocal phenomena may be quantitatively characterized?

Recently, it was found that the uncertainty principle governs the non-locality of quantum mechanics \cite{UR-determine}. In this work, we propose a method to further determine the different strengths of nonlocal correlations. First, in the framework of generalized uncertainty principle (GUP) \cite{GUP}, we demonstrate quantitatively that the uncertainty relation may govern the degrees of nonlocality ranging from the superquantum to quantum, and from the Bell local to non-steering, etc. Second, by attributing the uncertainty relation to the dependence relation of incompatible observables, new types of nonlocal phenomena that are fundamentally different from the BI violation are obtained. An example of the third order ``skewness nonlocality'' is constructed, and the Bell nonlocality turns out to be the second order ``variance nonlocality''. We provide as well concrete examples for the quantum contextuality which is found pertaining to the squares of commutators appearing in the forth-order dependence.

\section{The degrees of quantum nonlocality}

The fundamental postulates of QM tell that physical obervables may be represented by Hermitian matrices, and the measurement results of an observable can only be those eigenvalues of the Hermitian matrix. Two observables $X$ and $Y$ in $N$-dimensional representation may sum as $X+Y =Z$, and there exists the relation \cite{Horn-Conj}:
\begin{align}
\sum_{i=1}^{l}\alpha_i + \sum_{j=1}^l \beta_j \geq \sum_{k=1}^l\gamma_k\; , \; 1\leq l\leq N\; , \label{Horn-ineq}
\end{align}
where $\alpha_i$, $\beta_j$, and $\gamma_k$ are eigenvalues of $X$, $Y$, and $Z$, arranged in descending order. Note, the relation (\ref{Horn-ineq}) has various applications in quantum information sciences \cite{Horn-sep}. Suppose $Y$ and $Y'$ are two-dimensional observables with eigenvalues $\pm 1$, according to relation (\ref{Horn-ineq}) the summation $(Y-Y') + (Y+Y') = 2Y$ yields
\begin{align}
\alpha_1 + \beta_1 \geq \gamma_1= 2 \; . \label{YY'eigen}
\end{align}
Here $\alpha_1$, $\beta_1$, and $\gamma_1$ are the largest eigenvalues of $(Y-Y')$, $(Y+Y')$, and $2Y$, respectively. When $Y$ and $Y'$ are orthogonal qubit observables, e.g., Pauli matrices $Y=\sigma_x$ and $Y'=\sigma_z$, we have $\alpha_1=\beta_1=\sqrt{2}$ and then relation (\ref{YY'eigen}) exhibits the fact $\sqrt{2} +\sqrt{2} > 2$.

Now, let $y_i$ and $y_j'$ be the eigenvalues of $Y$ and $Y'$ with observing probabilities $p_{y_{i}}$ and $p_{y_j'}$, the expectation value of $Y+Y'$ can be expressed as
\begin{align}
\langle Y+Y' \rangle = ( \vec{y} \oplus \vec{y}\,') \cdot (\vec{p}_y \oplus \vec{p}_{y'}) \; ,
\end{align}
where $\vec{y}$ and $\vec{y}\,'$ are vectors composed of the eigenvalues, $\vec{p}_{y}$ and $\vec{p}_{y'}$ signify the corresponding probability distributions. For qubit observables, the following relation obviously holds by definition
\begin{align}
\langle Y+Y' \rangle \leq ( \vec{y} \oplus \vec{y}\,')^{\downarrow} \cdot \vec{s}^{\,\downarrow} \; . \label{maj-uncer-cons}
\end{align}
Here $\downarrow$ denotes that the components are rearranged in descending orders and $\vec{s}$ is the optimal bound for the majorization uncertainty relation $\vec{p}_{y} \oplus \vec{p}_{y'} \prec \vec{s}$ \cite{Optimal-UR}. Note, according to a recent study, the uncertainty relation can be interpreted as the dependence between different measurements \cite{GUP}. In this sense, the expectation value $\langle Y+Y'\rangle$ may reach $2$ when $Y$ and $Y'$ are independent observables with eigenvalues $\pm 1$. However, the uncertainty relation, i.e. $\vec{p}_{y} \oplus \vec{p}_{y'} \prec \vec{s}$\,, would limit the expectation value $\langle Y+Y'\rangle$ of a dependent pair of observables to be less than $2$ (see Appendix A).

With the above preparations we are ready to examine how quantum nonlocality emerges and behaves. Consider the correlation $E(X,Y)$ in CHSH inequality (\ref{CHSH-up2}), it may exhibit in LHVT and quantum theory respectively as
\begin{eqnarray}
\mathrm{LHVT:} & & E(X,Y) = \int \xi_{\lambda} A(\lambda,X)B(\lambda,Y) \, \mathrm{d}\lambda\; , \\
\mathrm{QM:} & & E(X,Y) =  \langle X\otimes Y\rangle\; .
\end{eqnarray}
Here $\xi_{\lambda}$ stands for the unknown distribution of some hidden variables $\lambda$, positive and normalized; $A(\lambda,X)$ and $B(\lambda,Y)$ are measurements performed by Alice and Bob respectively. In LHVT, the dichotomic functions $A(\lambda,X)$ and $B(\lambda,Y)$ are given the values of $\pm 1$, and are determined jointly by $\lambda$ and the observables, $X$ and $Y$.

\subsection{The LHVT correlations}

In LHVT, obviously the following inequality holds:
\begin{align}
-2 \leq &  A(\lambda,X)[B(\lambda,Y)-B(\lambda,Y')] +  A(\lambda,X')[B(\lambda,Y)+B(\lambda,Y')] \leq 2\; . \label{CHSH-LHVT}
\end{align}
The lower and upper bounds $\pm 2$ are obtained from the following arguments: 1. The values of $A(\lambda, X)$ and $A(\lambda, X')$ are independent and both can be $\pm 1$; 2. The values of $B(\lambda, Y)$ and $B(\lambda, Y')$ are also independent, while the combination of $B(\lambda,Y)-B(\lambda,Y')$ and $B(\lambda,Y)+B(\lambda,Y')$ falls in the scope of $[-2,2]$. Therefore, after integrating over the distribution $\xi_{\lambda}$, one can readily find the well-known CHSH inequality \cite{CHSH}
\begin{align}
-2 \leq & E(X,Y) -E(X,Y') + E(X',Y)+E(X',Y') \leq 2 \;.
\label{chsh}
\end{align}

\subsection{The non-steerable correlation}

For the steerability, a kind of quantum nonlocal correlation, if $A$(lice) cannot steer $B$(ob), then the hidden state $\sigma^{(\lambda)}$ of the $d$-dimensional
quantum system $B$, on condition of measurement result $i$ of any observable $X$, may be represented by the assemblage defined as a set of $d_B\times d_B$ Hermitian matrices \cite{assemblage-1}
\begin{align}
\sigma_{i|x} = \sum_{\lambda} \xi_{\lambda}\, p_{i}^{(\lambda)}(x) \sigma^{(\lambda)}\; . \label{Assemblages-x}
\end{align}
Here the probability distribution $p_i^{(\lambda)}(x)$ is normalized $\sum_{i}p_i^{(\lambda)}(x)=1$. Evaluate the correlations in (\ref{CHSH-up2}) by means of the above assemblage, the two terms on $B$ sector in Eq. (\ref{CHSH-LHVT}) turn out to be:
\begin{align}
B(\lambda,Y)-B(\lambda,Y') = \mathrm{Tr}[\sigma^{(\lambda)}(Y-Y')] \; , \label{CHSH-NSteering1} \\
B(\lambda,Y)+B(\lambda,Y') = \mathrm{Tr}[\sigma^{(\lambda)}(Y+Y')] \;.  \label{CHSH-NSteering2}
\end{align}
One may notice that: 1. The values of $A(\lambda, X)$ and $A(\lambda, X')$ remain independent and both can be $\pm 1$; 2.
$B(\lambda, Y)$ and $B(\lambda, Y')$ are not independent anymore, due to the uncertainty relation imposed on $\sigma^{(\lambda)}$ \cite{QMS}; 3. The uncertainty relation in form of equation (\ref{maj-uncer-cons}) constrains the magnitudes of (\ref{CHSH-NSteering1}) and (\ref{CHSH-NSteering2}) to be less than $\sqrt{2}$. And furthermore we have (see Appendix A for details)
\begin{align}
 \left[ E(X,Y) -E(X,Y') \right]^2 +
 \left[ E(X',Y)+E(X',Y')\right]^2 \leq 2 \;.
\label{sinequality}
\end{align}
Note that the relation (\ref{sinequality}) is a generally established condition for any non-steerable correlation, which complies with (\ref{chsh}).

\subsection{The quantum correlation}

To reveal the strength of QM correlation, following we construct a toy model in bipartite system, in which in lieu of assemblage (\ref{Assemblages-x}) we define
\begin{align}
\sigma_{i_1|x} & = \sum_{\lambda} \xi_{\lambda} \,  p_{i_1}^{(\lambda)}(x) \sigma^{(\lambda)}(x)  \; , \\
\sigma_{i_2|x'} & = \sum_{\lambda} \xi_{\lambda} \, p_{i_2}^{(\lambda)}(x') \sigma^{(\lambda)}(x') \;.
\end{align}
Note, different from $\sigma^{(\lambda)}$ in assemblage (\ref{Assemblages-x}), here the hidden states $\sigma^{(\lambda)}(x)$ and $\sigma^{(\lambda)}(x')$ rely on the measurements $X$ and $X'$ respectively, and are mutually independent. Then we readily get
\begin{align}
B(\lambda,Y)-B(\lambda,Y') = \mathrm{Tr}[\sigma^{(\lambda)}(x)(Y-Y')] \; , \label{CHSH-QM1}\\
B(\lambda,Y)+B(\lambda,Y') = \mathrm{Tr}[\sigma^{(\lambda)}(x')(Y+Y')] \; , \label{CHSH-QM2}
\end{align}
and may have the following observations: 1. $A(\lambda, X)$ and $A(\lambda, X')$ remain to be independent and both can achieve $\pm 1$; 2. $B(\lambda, Y)$ and $B(\lambda, Y')$ are interrelated on each other according to the uncertainty relation \cite{QMS}; 3. Unlike (\ref{CHSH-NSteering1}) and (\ref{CHSH-NSteering2}), equations (\ref{CHSH-QM1}) and (\ref{CHSH-QM2}) are independent with each other and may reach the corresponding maxima of $2\cos\frac{\theta}{2}$ and $2\sin\frac{\theta}{2}$ respectively (see Appendix A for the arguments). Hence we have
\begin{align}
[E(X,Y) -E(X,Y')]^2 +  [E(X',Y)+E(X',Y')]^2 \leq 4 \; . \label{CHSH-QMX}
\end{align}
Consider the inequality $(a+b)^2 \leq 2(a^2+b^2)$, one then notices that the equation (\ref{CHSH-QMX}) breaks the non-steerable condition (\ref{sinequality}) and CHSH inequality (\ref{chsh}), which may be regarded as a corollary of relations (\ref{YY'eigen}) and (\ref{maj-uncer-cons}).

\begin{figure}[t]\centering
\scalebox{0.6}{\includegraphics{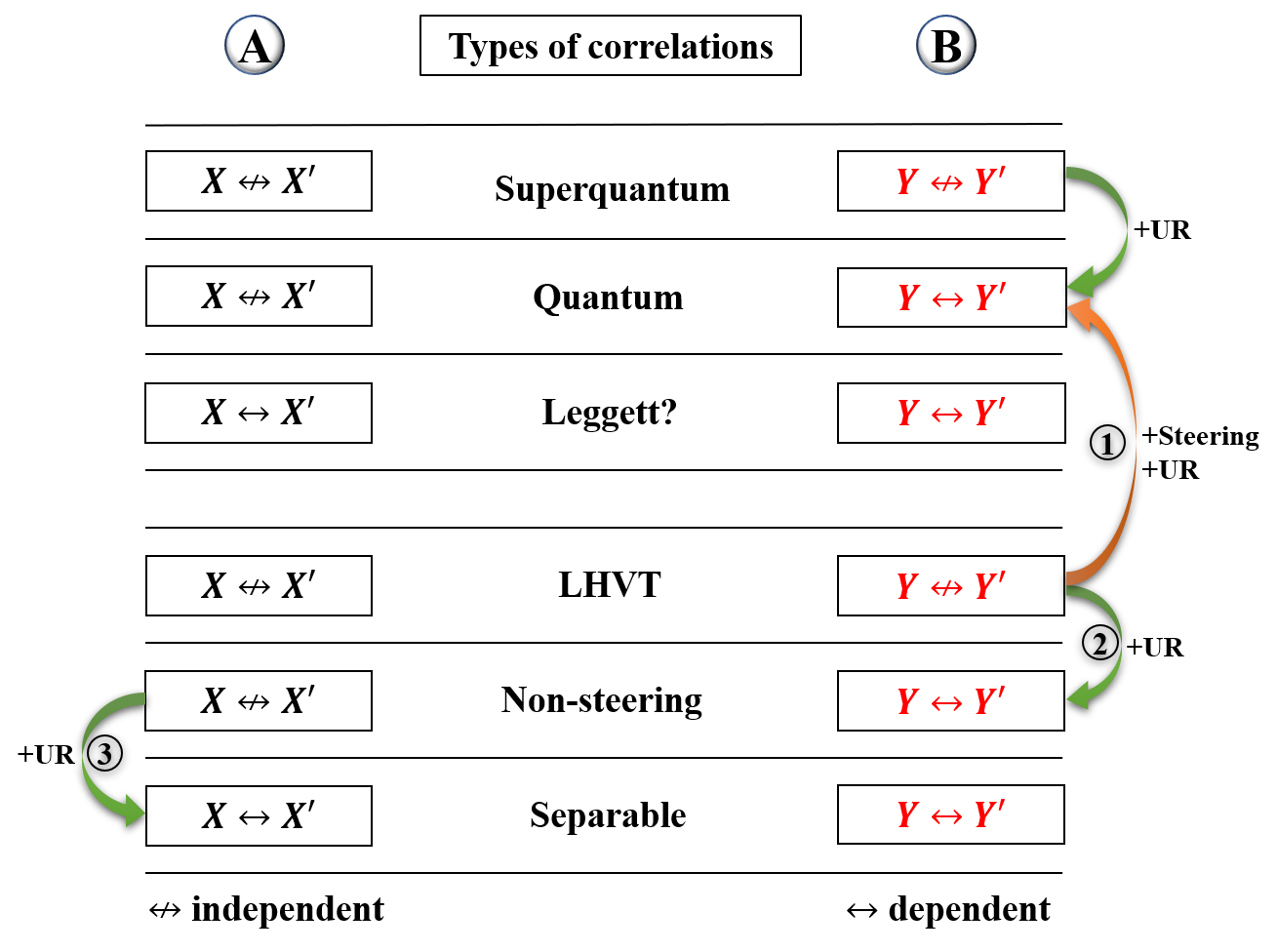}}
\caption{{\bf Various degrees of nonlocality}. The connections between different types of nonlocality are signified with arrows. Note, the relationship between Leggett model and other types of nonlocalities is still unclear. } \label{Figure-nonlocal}
\end{figure}

\subsection{The superquantum correlation}

Now we make a further assumption about the quantum mechanical results of equations (\ref{CHSH-QM1}) and (\ref{CHSH-QM2}): Let $Y$ and $Y'$ be independent observables, i.e., there is no uncertainty relation constraining them. Therefore, as discussed below (\ref{maj-uncer-cons}), we certainly have
\begin{align}
-4 \leq & E(X,Y) -E(X,Y') + E(X',Y)+E(X',Y') \leq 4 \;,
\end{align}
which gives a more broad range for correlations then CHSH (\ref{chsh}). We may think it as a kind of correlation beyond quantum mechanics, say superquantum correlation. In Figure \ref{Figure-nonlocal} different types of nonlocality of various models are presented, among them the connections between some types of nonlocalities had been investigated: \ding{192} is studied in Ref. \cite{UR-determine}; \ding{193} and \ding{194} are studied in Ref. \cite{QMS}. Note, whether the Leggett model \cite{Leggett-NHVT} could be assigned to the nonlocal pattern in Figure \ref{Figure-nonlocal} or not remains to be an interesting and open question. Next, we shall manifest how the nonlocal phenomenon behaves while higher order dependences are taken into account.

\section{Various quantum nonlocalities}

Even within the regime of quantum mechanics there are different tiers of nonlocality, which fortunately can be distinguished by the generalized quantum uncertainty principle, developed in Ref.\cite{GUP}. According to it, the uncertainty relation may be expanded in terms of cumulants, each corresponding to a certain strength of nonlocality. Here, in this work we find the different orders of nonlocality can be employed to characterize the various quantum correlations.

Given a random variable $X$, the moment generating function takes the following form
\begin{align}
\langle e^{sX}\rangle =\sum_{n=0}^{\infty}\langle X^n\rangle \frac{s^n}{n!} \; ,\; s\in \mathbb{C}\; . \label{Moment-Gen}
\end{align}
Here $\langle X\rangle$ means the expectation value of a variable $X$ and the parameter $s$ is a complex number. The logarithm of equation (\ref{Moment-Gen}) generates the cumulants \cite{Stuart2010}, that is
\begin{align}
K(sX) & \equiv \log(\langle e^{sX}\rangle)= \log\left(1+s\langle X\rangle + \frac{s^2}{2!}\langle X^2\rangle + \frac{s^3}{3!}\langle X^3\rangle + \cdots \right) \nonumber \\
& = \sum_{m=1}^{\infty} \frac{s^m}{m!}\kappa_m(X) \; , \label{Kx-Expansion}
\end{align}
where the sum runs over a power series of $s$ whose coefficients $\kappa_m(X)$ are called the $m$th order cumulant.

According to Ref. \cite{GUP}, for arbitrary observables $X$ and $Y$, there exists a generalized uncertainty relation
\begin{align}
K[(s+s^*)X] + K[(t+t^*)Y] \geq K(Z_{st}) +  K^*(Z_{st})  \; , \; s, t \in \mathbb{C} \; . \label{K-uncertainty}
\end{align}
Here $K(\cdot)$ signifies the generating function of cumulants defined in equation (\ref{Kx-Expansion}); * means the complex conjugation; $Z_{st}= \log(e^{sX}e^{tY}) = Z_{1}+ Z_{11}+\cdots$ is defined as
\begin{align}
Z_1 = sX +tY\; , \; Z_{11} = \frac{1}{2}[sX,tY]\;, \cdots \; ,\label{Haus-Z}
\end{align}
in light of the well-known Baker-Campbell-Hausdorff (BCH) formula.

\subsection{The second order: commutators and Bell nonlocality}

For any bipartite system, a joint operation of measurement may be expressed as $\mathcal{S} = \sum_{i,j}m_{ij}X_i\otimes Y_j$ with $m_{ij} \in \mathds{R}$. Note, the $n$th order cumulant $\kappa_{n}(\mathcal{S})$ exists, given the $n$th and lower orders of moments of an observable exist \cite{Stuart2010}.  For illustration, we consider a typical representative joint observable of the bipartite qubit system
\begin{align}
S \equiv X\otimes Y - X\otimes Y' + X' \otimes Y + X' \otimes Y' \label{CHSH-S-operator}
\end{align}
with local representations
\begin{align}
X= \sigma_x\;,\; X'=\sigma_y \; ,\; Y = \cos\theta\sigma_x+\sin\theta\sigma_y\; ,\; Y' = -\sin\theta\sigma_x+\cos\theta \sigma_y\; . \label{XYtheta}
\end{align}
Here $X$ and $X'$ are orthogonal, and so do the $Y$ and $Y'$.

The second-order cumulant is the variance $\kappa_{2}(S) \equiv \langle S^2\rangle - \langle S\rangle^2$. For LHVT, the cumulant $0\leq \kappa_{2}(S)\leq 4$ (see details in Appendix B), and we have
\begin{proposition}
A bipartite system possesses the second-order nonlocality if the following Bell inequality is violated
\begin{align}
\kappa_2(S)\geq 0 \Rightarrow |\langle S\rangle | \leq 2 \; , \label{kappa2great0}
\end{align}
which is in fact the CHSH inequality $\left| E(X,Y) -E(X,Y') + E(X',Y) + E(X',Y') \right|\leq 2$.
\end{proposition}
The key point in deriving equation (\ref{kappa2great0}) is the evaluation of $ S^2= 4 I \otimes I  + [X,X'] \otimes [Y,Y'] $. The expectation values of commutators are supposed to be zero for LHVT \cite{Fine1, Fine4Landau}, and we readily arrive at the CHSH inequality $|\langle S\rangle| \leq 2$, see Figure \ref{Figure-kappa3}(a).

\subsection{The third order: the skewness of non-classical correlation}

\begin{figure}\centering
\scalebox{0.37}{\includegraphics{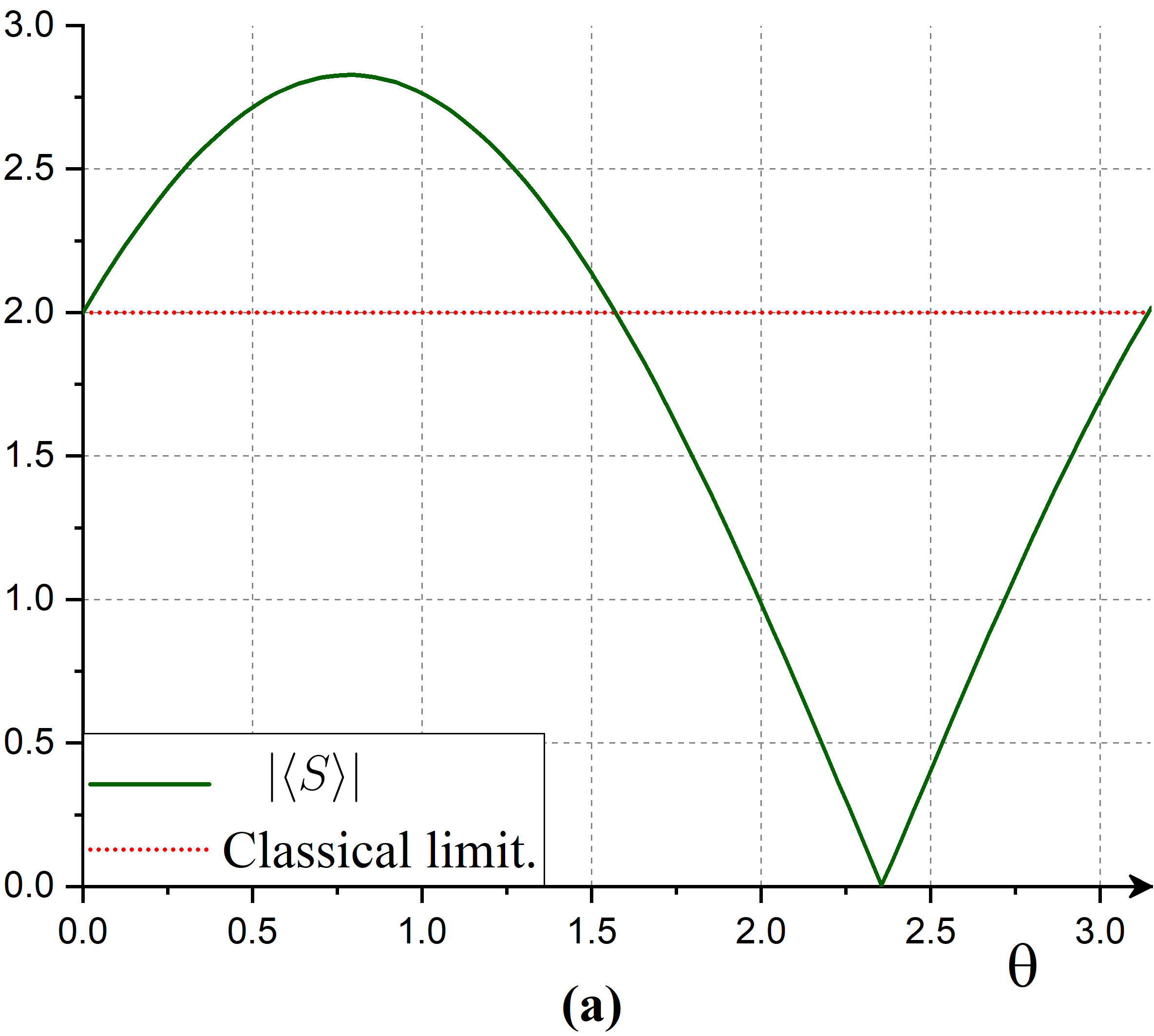}}\;
\scalebox{0.37}{\includegraphics{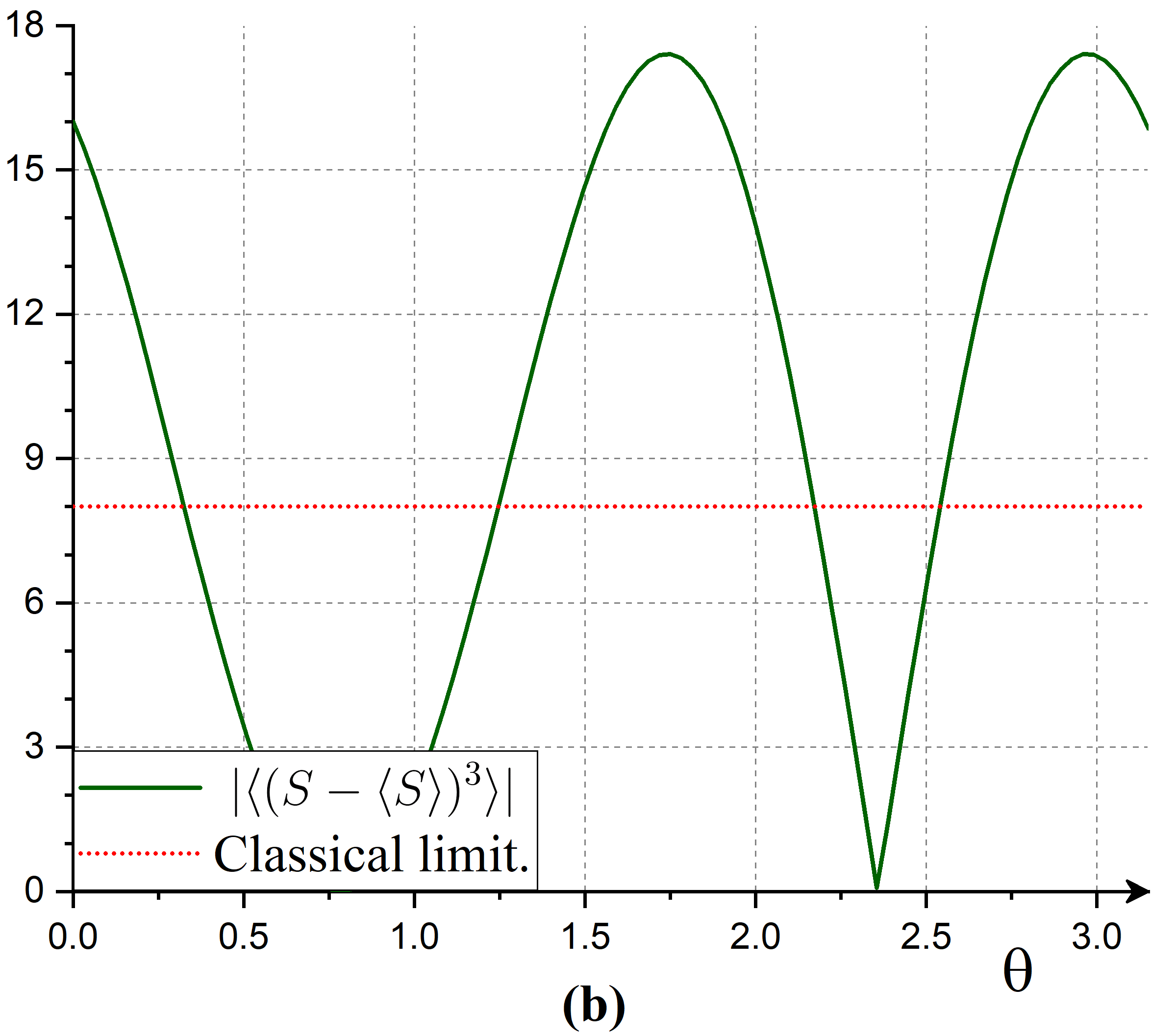}}
\caption{{\bf The Bell nonlocality and the skewness nonlocality.} In the spin singlet state: (a) The quantum prediction of $|\langle S\rangle|$ may reach the value of $2\sqrt{2}$ which violates the classical limit of 2; (b) The quantum prediction of skewness $|\kappa_3(S)| = |\langle (S -\langle S\rangle)^3\rangle| $ may reach a pretty high value of $64\sqrt{6}/9$ which evidently violates the classical limit of $8$.} \label{Figure-kappa3}
\end{figure}
The third order cumulant names the skewness, i.e. $\kappa_3(S) \equiv \langle S^3\rangle -3\langle S^2\rangle \langle S\rangle + 2\langle S\rangle^3$. Considering that in LHVT, for a typical observable with expectation value satisfying $-2\leq \langle S\rangle \leq 2$, the cumulant $|\kappa_3(S)|$ in classical statistics has the limit of $8$ \cite{Central-Moments}, we then have:
\begin{proposition}
A bipartite system contains the third-order nonlocality if the following ``skewness'' inequality is violated
\begin{align}
\left| \kappa_3(S) \right| = \left| \langle \left( S-\langle S\rangle \right)^3\rangle \right|\leq  8\; . \label{Kappa3}
\end{align}
Here $S$ is defined as in equation (\ref{CHSH-S-operator}).
\end{proposition}
The key point in deriving equation (\ref{Kappa3}) is the evaluation of the high order commutators like $[[X, X'], X]$, whose expectation values are zeros in the joint distribution model of LHVT \cite{Fine1} (see Appendix B and C). The QM prediction for relation (\ref{Kappa3}) in spin singlet state is plotted as Figure \ref{Figure-kappa3}(b).

\subsection{The forth order: the commutator squares and contextuality}

\begin{figure}\centering
\scalebox{0.45}{\includegraphics{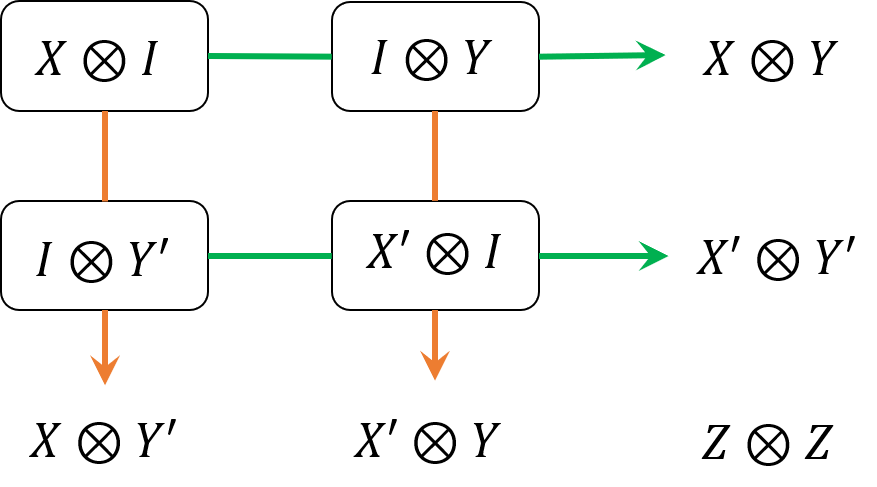}}
\caption{{\bf The contextuality of two spin-1/2 particles.} The boxed operators in each column and row are commutative and thus simultaneously measurable. The four boxed quantities, when multiplying in different orders, may lead to some contradictive results. } \label{Figure-context}
\end{figure}

In the forth order cumulant
\begin{align}
S^4  = 16 I\otimes I+ [X,X']^2 \otimes [Y,Y']^2 + 8 [X,X'] \otimes [Y,Y']  \; ,
\label{forthcumulant}
\end{align}
a new type of operator appears, that is the second term on the right hand side of the equation. With the operator choice in  (\ref{XYtheta}), one can readily find $[X,X']^2 \otimes [Y,Y']^2=16I\otimes I$ and then the (\ref{forthcumulant}) turns to
\begin{align}
S^4 = 32 I\otimes I + 8 [X,X'] \otimes [Y,Y'] \; .
\end{align}
Note, in LHVT the expectation value of nontrivial commutator is not well defined. For instance the observable $L_z^2$ may have nontrivial expectation value, while $(i[L_x,L_y])^2$ is identically zero in any joint distribution model of LHVT. We shall show below how the commutator squared in the fourth cumulant (\ref{forthcumulant}) implies for the KS contextuality.

Consider the KS contextuality of two spin-1/2 particles given in Ref. \cite{NDMerminRMP}, the measurements in each row and column of Figure \ref{Figure-context} are commutable, e.g., the first row $\{X\otimes I,\; I\otimes Y,\; X\otimes Y\}$, where $X$, $X'$, $Y$, and $Y'$ are defined in (\ref{XYtheta}) and $Z=\sigma_z$. Multiplying the observables in boxes of Figure \ref{Figure-context} in rows we have $XX'\otimes YY'=R$, while in columns we get $XX'\otimes Y'Y=C$. Since the values assigning to $R$ and $C$ should be the same in classical point of view, their product is then a square number and positive. While in QM, the following expression is apparently negative due to commutator squared
\begin{align}
RC=(XX'\otimes YY') (XX'\otimes Y'Y) = \frac{1}{4}[X,X']^2\otimes I \; , \label{2-qubits-context}
\end{align}
where relations $XX'= \frac{1}{2}([X, X']+\{X, X'\})=[X,X']/2$ and $Y^2=Y'^2=I$ are employed. Taking into account what discussed in Sec 3.1, we may make the following conjecture:
\begin{conjecture}
The BI violation is related to the nontrivial expectation value of commutators, while the contextuality is related to the nontrivial expectation values of the commutator squares (or higher powers). \label{Conject-1}
\end{conjecture}

From Conjecture \ref{Conject-1} we notice that the KS contextuality \cite{KS-noncotext1} may relate to the squares of commutators (details given in Appendix D). Though to establish an explicit and quantitative relation between contextuality and powers of commutators still needs more works, it is yet reasonable to believe that the correspondence of different nonlocal phenomena to dependent orders of incompatible observables should exist.

\section{Conclusions}

We demonstrate in this work that one may characterize the degree of nonlocality, from superquantum to classical, by exploiting the generalized uncertainty relation. It is found that in a micro world where entangled states exist but without uncertainty constraint, the magnitude of correlations constrained by CHSH inequality may reach maximally $4$. However, the operators in QM satisfy the uncertainty relation, which constrains the CHSH inequality to an upper bound of $2\sqrt{2}$. For classically correlated real observables, which are in separable states and has no uncertainty relation, the correlations of LHVT in CHSH inequality have an upper limit of $2$. Moreover, novel steering and separability criteria are obtained in addition to the above results.

In the second part of this paper, we signify different strengths of non-local correlations in quantum physics. The higher order dependence of observables existing in the generalized uncertainty relation found corresponds to the higher order non-classical phenomenon. By dint of an explicit example of ``skewness nonlocality'', the Bell nonlocality shown behaves as the ``variance nonlocality''. Considering commutator squares, the quantum contextuality is thought a non-classical phenomenon lying in the forth order dependence. Remarkably, we notice that the square of commutator had already found applications in describing quantum chaos in many body systems \cite{QChaos1}. It is expected that the higher order dependence may unveil the yet unknown non-classical phenomena and have some unique applications in quantum information, quantum computation, and quantum many-body system.

\section*{Acknowledgements}
\noindent
This work was supported in part by the National Natural Science Foundation of China(NSFC) under the Grants 11975236 and 12235008, and by the University of Chinese Academy of Sciences.

\section*{Author Contributions}
\noindent
All authors have equally contributed to the main result, the examples and the writing. All authors have given approval for the final version of the manuscript.

\section*{Competing Interests}
\noindent
The authors declare no competing interests.

\newpage
\setcounter{figure}{0}
\renewcommand{\thefigure}{S\arabic{figure}}
\setcounter{equation}{0}
\renewcommand\theequation{S\arabic{equation}}
\setcounter{theorem}{0}
\renewcommand{\thetheorem}{S\arabic{theorem}}
\setcounter{observation}{0}
\renewcommand{\theobservation}{S\arabic{observation}}
\setcounter{proposition}{0}
\renewcommand{\theproposition}{S\arabic{proposition}}
\setcounter{lemma}{0}
\renewcommand{\thelemma}{S\arabic{lemma}}
\setcounter{corollary}{0}
\renewcommand{\thecorollary}{S\arabic{corollary}}
\setcounter{section}{0}
\renewcommand{\thesection}{S\arabic{section}}

\appendix{\bf \Huge Appendix}

\section{The constraints from the direct sum majorization uncertainty relation}

Consider the following qubit observables
\begin{align}
Y = \sigma_z = \begin{pmatrix}
1 & 0 \\
0 & -1
\end{pmatrix} \; ,\; Y' = \cos\theta\sigma_z + \sin\theta \sigma_x =\begin{pmatrix}
\cos\theta & \sin\theta \\
\sin\theta & -\cos\theta
\end{pmatrix} \;
\end{align}
with $\theta \in [0,\pi/2]$, there exists the following direct sum majorization uncertainty relation \cite{S-opt-maj}
\begin{align}
\vec{p}_y\oplus \vec{p}_{y'} \prec \vec{s} =
\begin{pmatrix}
1 \\
\cos\frac{\theta}{2} \\
1-\cos\frac{\theta}{2} \\
0
\end{pmatrix}\;.
\end{align}
The expectation values of $Y+Y'$ and $Y-Y'$ have the following constraints
\begin{align}
\langle \psi|Y+Y'|\psi \rangle & = (\vec{y} \oplus \vec{y}\,')^{\downarrow} \cdot (\vec{p}_y\oplus \vec{p}_{y'}) \leq 2\cos\frac{\theta}{2} \; , \\
\langle \phi|Y-Y'|\phi \rangle & = (\vec{y} \oplus -\vec{y}\,')^{\downarrow} \cdot (\vec{p}_y\oplus \vec{p}_{-y'}) \leq 2\sin\frac{\theta}{2} \; ,
\end{align}
where the eigenvalue vectors satisfy $(\vec{y} \oplus \vec{y}\,')^{\downarrow}=(\vec{y} \oplus -\vec{y}\,')^{\downarrow}=(1,1,-1,-1)^{\mathrm{T}}$.

On the other hand, for orthogonal observables $Y$ and $Y'$ that $\theta=\pi/2$, we would have
\begin{align}
\mathrm{Tr}[\rho (Y+Y')] = a \sqrt{2} \; ,\;
\mathrm{Tr}[\rho (Y-Y')] = a' \sqrt{2} \;.
\end{align}
Here density matrix $\rho = \frac{\mathds{1}}{2} + \frac{1}{2}\vec{r} \cdot\vec{\sigma}$; $a^2+a'^2\leq 1$ with $a$ and $a'$ being about the angles between the Bloch vectors of $\vec{r}$ and $Y+Y'$ and $Y-Y'$, respectively. Note, the orthogonal situation of $Y$ and $Y'$ pair gives us the strongest constraint on (S5).

\section{The second-order cumulant: variance nonlocality}

\begin{figure}[t]\centering
\scalebox{0.45}{\includegraphics{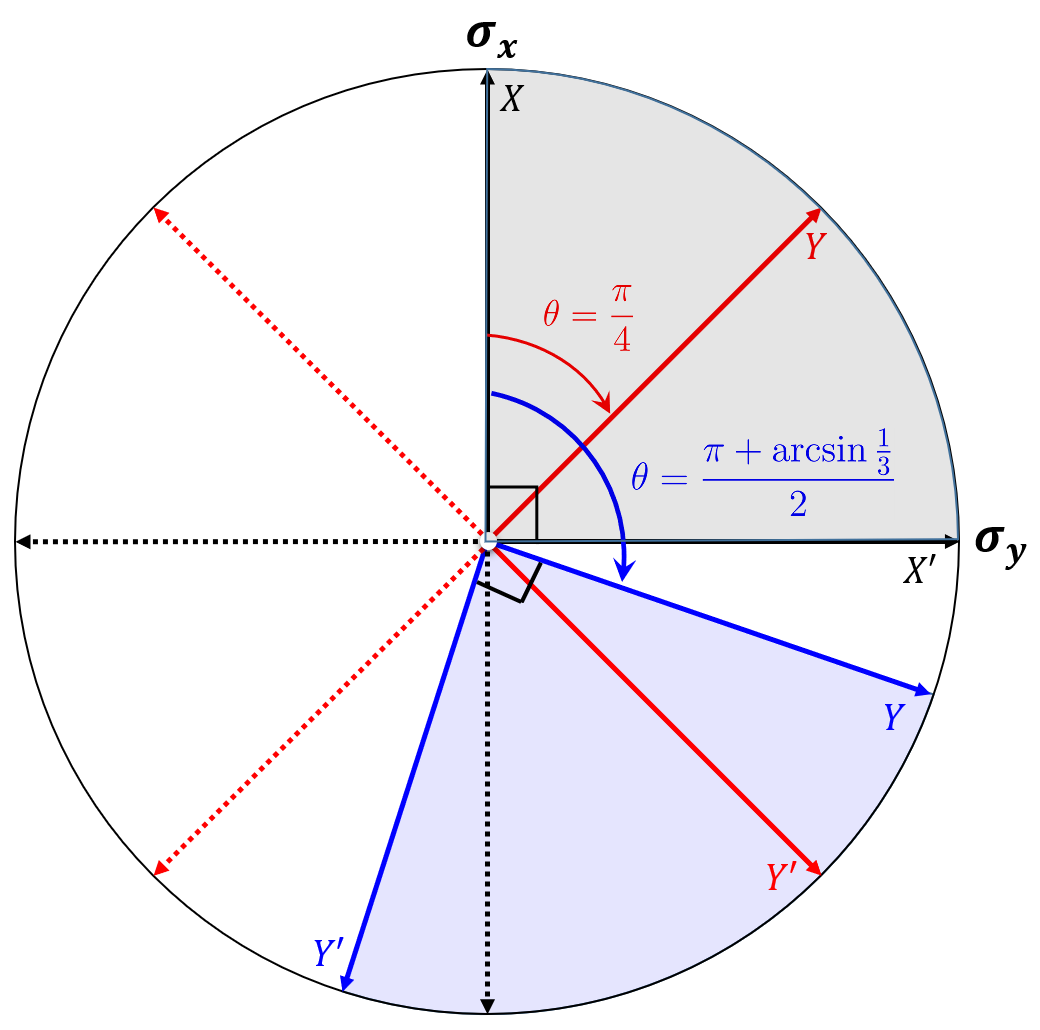}}
\caption{{\bf A typical configuration of observables $X$-$X'$ on $A$ and $Y$-$Y'$ on $B$.} Here $X$ and $X'$ in Bloch vector form represent orthogonal spin observables, so are the $Y$ and $Y'$. The relative angle between $X$ and $Y$ is $\theta \in [0,\pi]$.  } \label{S-Figure-XY}
\end{figure}

According to the definition in the main text, the operator $S$ for qubit observables is
\begin{align}
S \equiv X\otimes Y - X\otimes Y' + X' \otimes Y + X' \otimes Y' \; , \label{S-S-operator}
\end{align}
where $X= \sigma_x$, $X'=\sigma_y$, $Y = \cos\theta\sigma_x+\sin\theta\sigma_y$, and $Y' = -\sin\theta\sigma_x+\cos\theta \sigma_y$, see Figure \ref{S-Figure-XY}. The second-order cumulant is defined as \cite{S-GUP}
\begin{align}
\kappa_2(S) & = \langle S^2\rangle - \langle S\rangle^2 \; .
\end{align}
Obviously $\kappa_2(S)$ is just the variance and
\begin{align}
S^2 = & 4 (\mathds{1} \otimes \mathds{1}) + [X,X'] \otimes [Y,Y'] \; . \label{S-S2}
\end{align}
For LHVT which gives the CHSH type inequalities, there exists a joint distribution model for the four bivalent observables of $X$, $X'$, $Y$, and $Y'$, known as the Fine's theorem \cite{S-Fine1} (see Refs. \cite{S-Fine2,S-Fine3} for more discussions on the Fine's theorem). Of the joint distribution model in LHVT, the expectation values of commutators shall also be zeros, say
\begin{align}
\langle [X,X']\otimes [Y,Y']\rangle = 0\;.
\end{align}
The LHVT predictions for the cumulants are therefore
\begin{align}
\kappa_2(S) & = 4 - \langle S\rangle^2 \; . \label{S-k-variance}
\end{align}
Since the classical statistical lower bound for the variance $\kappa_2(S)$ is $0$, we then have
\begin{align}
0\leq \langle S\rangle ^2 \leq 4 \; . \label{S-K-var2}
\end{align}
Equation (\ref{S-K-var2}) provides a LHVT prediction for the correlation functions of $\langle S\rangle = E(X,Y) -E(X,Y') +E(X',Y) + E(X',Y')$.

In QM, for orthogonal spin observables $X \bot X'$ ($Y\bot Y'$, see Figure \ref{S-Figure-XY}) taken as in (\ref{S-S-operator}), we may get
\begin{align}
S^2 & = 4 + 2 i\sigma_z \otimes 2i\sigma_z \; .
\end{align}
Here the relations $[X,X']=[Y,Y']=2i\sigma_z$ are applied to equation (\ref{S-S2}). For spin singlet state
\begin{align}
|\psi\rangle = \frac{1}{\sqrt{2}} (|+-\rangle -|-+\rangle) \; ,
\end{align}
the expectation value of equation (\ref{S-S2}) becomes
\begin{align}
\langle S^2 \rangle = 8 \ . \label{exps2}
\end{align}
Hence the second order variance in QM turns out to be
\begin{align}
\kappa_2(S) & = 8 -\langle S\rangle^2 \in [0,8]\; .
\end{align}
Clearly the value of $\kappa_2(S)$ in QM violates the LHVT prediction of (\ref{S-K-var2}).

\section{The third order: Skewness nonlocality}

For the operator $S$, the third order cumulant is defined as \cite{S-GUP}
\begin{align}
\kappa_3(S) & = \langle S^3\rangle -3\langle S^2\rangle \langle S\rangle + 2\langle S\rangle^3 \;.
\end{align}
The 3rd power of observable $S$ writes
\begin{align}
S^3 = & 4 (X\otimes Y - X\otimes Y' + X'\otimes Y + X'\otimes Y') + \nonumber \\ & \frac{1}{4}[[X,X'],X] \otimes[[Y,Y'],Y] - \frac{1}{4}[[X,X'],X] \otimes [[Y,Y'],Y']+ \nonumber \\ & \frac{1}{4}[[X,X'],X'] \otimes [[Y,Y'],Y] + \frac{1}{4} [[X,X'],X'] \otimes [[Y,Y'],Y'] \; . \label{S-S3}
\end{align}
In the joint distribution model of LHVT, the classical statistical boundary for skewness is $|\kappa_3(S)|\leq 8$ for the variable with expectation value $\langle S\rangle \in [-2,2]$ \cite{Cent-Moments}, and therefore
\begin{align}
\left| \langle S^3\rangle -3\langle S^2\rangle \langle S\rangle + 2\langle S\rangle^3 \right| \leq 8 \; . \label{S-K-Skew2}
\end{align}
Relation (\ref{S-K-Skew2}) is the LHVT prediction of the correlation functions in $\langle S\rangle = E(X,Y) -E(X,Y') +E(X',Y) + E(X',Y')$.

In QM, for orthogonal spin observables $X \bot X'$ ($Y\bot Y'$, see Figure \ref{S-Figure-XY}) chosen in equation (\ref{S-S-operator}), it can be shown that
\begin{align}
S^3 & = 8 (X\otimes Y - X\otimes Y' + X'\otimes Y + X'\otimes Y')\; .
\end{align}
Here the relation $[[X,X'],X]=-4X'$ is employed in the calculation of (\ref{S-S3}), and similarly other commutators in it are evaluated. For spin singlet state
\begin{align}
|\psi\rangle = \frac{1}{\sqrt{2}} (|+-\rangle -|-+\rangle) \; ,
\end{align}
the expectation value of $S^3$ in (\ref{S-S3}) becomes
\begin{align}
\langle S^3\rangle = 8 \langle S\rangle\; . \label{exps3}
\end{align}
Hence the 3rd cumulant in QM turns out to be
\begin{align}
\kappa_3(S) & = 8 \langle S\rangle - 24 \langle S\rangle + 2\langle S\rangle^3  = 2\left(\langle S\rangle^3 - 8 \langle S\rangle \right) \in [-\frac{64\sqrt{6}}{9}, \frac{64\sqrt{6}}{9}] \;.
\end{align}
Clearly the magnitude of $\kappa_3(S)$ in QM is greatly over the LHVT prediction in relation (\ref{S-K-Skew2}).

\section{The forth order: Contextuality}

\begin{figure}\centering
\scalebox{0.5}{\includegraphics{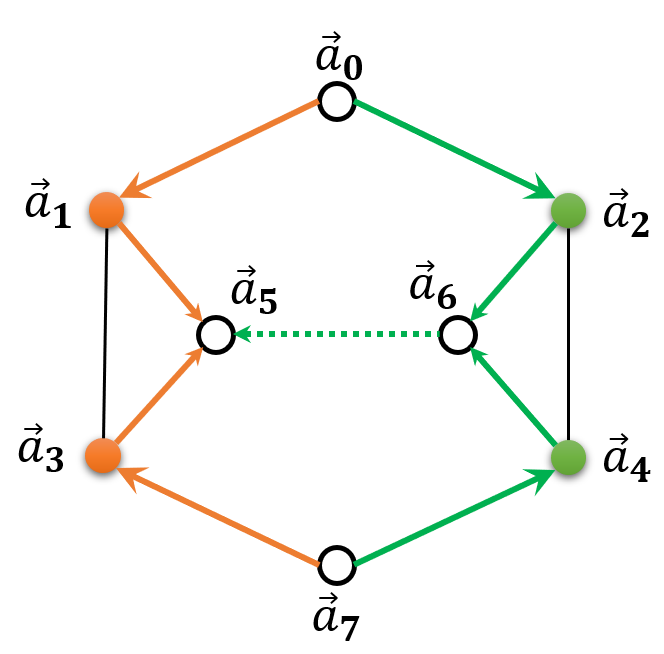}}
\caption{{\bf A contextuality structure in spin-1 system \cite{S-KS-noncotext1}.} The connected vectors are orthogonal: $\vec{a}_0\perp \vec{a}_1$, $\vec{a}_0\perp \vec{a}_2$, $\vec{a}_7 \perp \vec{a}_3$, $\vec{a}_7 \perp \vec{a}_4$, $\vec{a}_5 \perp \vec{a}_6$, and $\vec{a}_1 \perp \vec{a}_3 \perp \vec{a}_5$, $\vec{a}_2 \perp \vec{a}_4 \perp \vec{a}_6$. }\label{S-Figure-KS8}
\end{figure}

The KS contextuality was originally demonstrated using 117 real directions for spin-1 particle with a group of octet-vector sets $\{\vec{a}_i| i=0,1,\cdots,7\}$ constructed, as shown in Figure \ref{S-Figure-KS8} \cite{S-KS-noncotext1}, where vector rays in two directly jointed vertices are orthogonal. For example, on the left half the figure, we have $\vec{a}_0 \perp \vec{a}_1$, $\vec{a}_1 \perp \vec{a}_5 \perp \vec{a}_3$, and $\vec{a}_3 \perp \vec{a}_7$ (Similarly for the right half of the figure). In the case of spin-1 operators $L_i \equiv \vec{L}\cdot \vec{a}_i$ in real space, $[L_i^2,L_j^2]=0$ if $\vec{a}_i\perp \vec{a}_j$. For spin-1 system of $L^2=2$, the following conclusions hold \cite{S-KS-noncotext1}:
\begin{enumerate}
\item For any orthogonal frame, i.e., $\vec{a}_i\perp \vec{a}_j \perp \vec{a}_k$, $L_{i,j,k}^2$ take 0 exactly once;
\item For any orthogonal pair $\vec{a}_i\perp \vec{a}_j$, $L^2_{i,j}$ take 0 at most once.
\end{enumerate}
An obvious contradiction appears in case we assign zeros to $L_0^2$ and $L_7^2$ simultaneously, which nevertheless may have nonzero probabilities in QM.

The above contradiction in fact has inherent connection with the commutators squared. For orthogonal bases $\{\vec{a}_1,\vec{a}_3,\vec{a}_5\}$ and $\{\vec{a}_2,\vec{a}_4,\vec{a}_6\}$, we have
\begin{align}
L_5^2 = \left( i[L_1,L_3] \right)^2 \; ,\; L_6^2 = \left( i[L_2,L_4] \right)^2 \; . \label{S-KS56}
\end{align}
In case $L_0^2=L_7^2=0$, the processes to assign values for the commutative pairs $\{L_1^2,L_3^2\}$ and $\{L_2^2,L_4^2\}$ may go along different routes, i.e., the orange and green arrows in Figure \ref{S-Figure-KS8}. In QM, the values of $L_5^2$ and $L_6^2$ depend on the routes as per the squares of commutators in equation (\ref{S-KS56}). However, there is no proper classical definition on how to assign values for observables that appear as commutators squared. Therefore, the evaluations of $L_5^2$ and $L_6^2$ in classical theory fail to reflect the route dependence due to the quantum commutators.

\section*{Data Availability}
\noindent
All relevant data used for Examples and Figs. are available from the authors.

\end{document}